# Jamming in sheared foams and emulsions, explained by critical instability of the films between neighboring bubbles and drops

N. D. Denkov,[1,*] S. Tcholakova,[1] K. Golemanov,[1] A. Lips[2]

[1]*Laboratory of Chemical Physics & Engineering, Faculty of Chemistry, Sofia University, Bulgaria*
[2]*Unilever Global Research Center, Trumbull, Connecticut 06611, USA*

Phenomenon of foam and emulsion jamming at low shear rates is explained by considering the dynamics of thinning of the transient films, formed between neighboring bubbles and drops. After gradually thinning down to a critical thickness, these films undergo instability transition and thin stepwise, forming the so-called "black films", which are only several nanometers thick and, thereby, lead to strong adhesion between the dispersed particles. Theoretical analysis shows that such film thickness instability occurs only if the contact time between the bubbles/drops in sheared foam/emulsion is sufficiently long, which corresponds to sufficiently low (critical) rate of shear. Explicit expression for this critical rate is proposed and compared to experimental data.

Non-homogeneous flow, often discussed in terms of "shear banding" or "jamming-unjamming transitions", attracted researchers' attention in several areas, because it appears as a generic phenomenon in various systems, such as glassy and granular materials, concentrated suspensions, foams, emulsions, and micellar solutions [1-10]. This phenomenon is still poorly understood and appropriate theoretical modeling, beyond the phenomenological description, is missing. Foams and emulsions seem particularly suitable for studying and modeling such non-homogenous flow and related phenomena, because the behavior of these systems is governed by a relatively well understood interplay of capillary effects and viscous friction in the films, formed between neighboring bubbles and drops [11-15]. This understanding provides the unique possibility for detailed theoretical modeling and experimental studies of these systems at the microstructural level (viz. at the level of single drops, bubbles, and films), which is impossible for the other systems of interest.

Recently, several systematic studies were performed [1-10] to clarify the main factors controlling jamming/unjamming transitions in foams and emulsions. Some of the conclusions, relevant to the current study are: (i) Jamming is observed at a certain "critical" shear rate. When this critical rate is reached from above, the bubbles/drops in the dispersion "stick" to each other, thus creating jammed zones. (ii) Critical shear rate depends on several factors, such as the drop and bubble size, volume fraction, and most importantly, on the interaction between dispersed particles. The effect of interparticle forces was convincingly demonstrated with moderately concentrated emulsions (drop volume fraction $\Phi = 0.73$), for which jamming transition was observed in the systems with attractive interdroplet forces only [10]. Until now these observations lack clear explanations and quantitative description.

The main purpose of the current letter is to demonstrate that jamming transitions in flowing foams and emulsions could be explained by considering the dynamics of thinning of the films formed between neighboring bubbles and drops. For brevity, in most of the consideration below we discuss explicitly only bubbles in foams. However, the analysis could be applied to concentrated emulsions containing micrometer-sized drops, provided that the appropriate system parameters are used in the calculations.

The following consideration is largely based on our recent study [11-12], where we analyzed the dynamics of film thinning in relation to viscous dissipation in steadily sheared foams and emulsions. As in [11-12] we consider the processes of film formation and thinning between bubbles, located in two neighboring planes of sheared foam. These planes are assumed to slide along each other with constant relative velocity, $u$ (see Fig. 1).

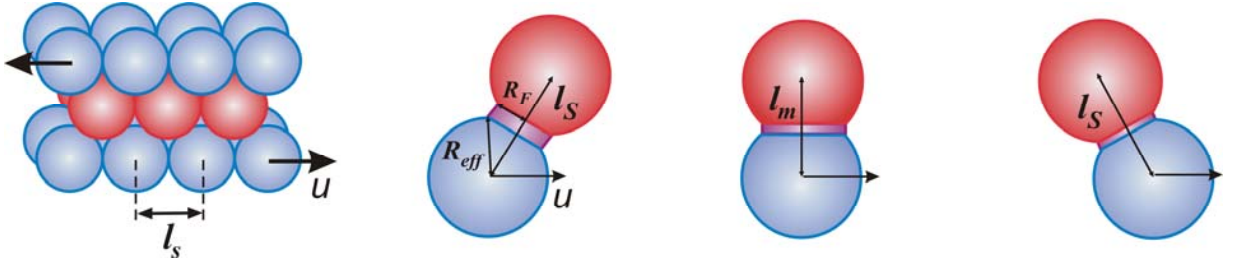

**FIG. 1.** Schematic presentation of the relative motion of neighboring planes of bubbles in sheared foam, and of the related process of film formation and thinning between two neighboring bubbles.

In such sheared foam, planar foam films with initial thickness, $h_0$, are formed, when the hydrodynamic pressure in the gap between the colliding bubbles becomes equal to the bubble capillary pressure, $P_d(h_0) = P_C$ [14-15]. From this pressure balance one can derive the following expression for the initial film thickness, $h_0 \approx (3\mu u/8\sigma)^{1/2} R_N$, where $\mu$ is dynamic viscosity of continuous phase, $\sigma$ is interfacial tension, and $R_N$ is radius of curvature of the bubble surface in the contact zone, just before the foam film formation [12,15]. As in Ref. [11], we assume that $R_N$ is approximately equal to the radius of curvature of the nodes in the static foam, $R_N = 2\sigma/P_C$, where $P_C(\Phi)$ is a known function of bubble volume fraction, $\Phi$ [13,16]. In the assumed bubble arrangement, the relative velocity of the bubbles, $u$, is related to the shear rate of the foam, $\dot{\gamma}$, by the expression, $\dot{\gamma} = 0.676 u \Phi^{1/3}/R_0$, and the capillary number is $Ca \equiv (\mu \dot{\gamma} R_0/\sigma)$.

The radius of the transient foam film, formed between two bubbles during their contact, $R_F(t)$, gradually increases from its initial value, $R_{F0} = (h_0 R_N)^{1/2}$, to a maximal value when the bubbles are closest to each other, and then decreases down to zero, as the bubbles separate dragged by the flow [11-12]. For all calculations involving $R_F(t)$, it is convenient to introduce an effective size of the bubbles, $R_{EFF} = (R_{FS}^2 + l_S^2/4)^{1/2}$, which is defined as the radius of a spherical surface with just one film (instead of the 12 films in the assumed fcc-structure) that has the same ratio $R_{FS}/l_S$, as the deformed polyhedral bubbles in the actual foam [12]. Here $R_{FS}(\Phi)$ is film radius and $l_S \approx 1.812 R_0/\Phi^{1/3}$ is center-to-center distance in the static non-sheared foam. Thus we replace the real polyhedral bubbles in the sheared foam by "imaginary" bubbles having just one foam film in the zone of contact, which allows us to use [12] the following approximate expression for the film radius, $R_F(t) = \left[R_{EFF}^2 - l(t)^2/4\right]^{1/2}$. Here $l(t) = \left[l_m^2 + \left(ut - \sqrt{l_0^2 - l_m^2}\right)^2\right]^{1/2}$ is the distance between bubble geometrical centers in sheared foam, $l_m = l_S \sqrt{3}/2$ is the minimal distance between bubble centers, and $l_0 = l(t=0)$ is the distance in the moment of film formation.

Next, the theoretical analysis shows [12] that the thickness of the formed film, $h(t)$, obeys the well-known Reynolds equation

$$(dh/dt) = -2\left[P_C - \Pi(h)\right]h^3 / 3\mu R_F^2 \qquad (1)$$

where $\Pi(h)$ is the disjoining pressure, which accounts for the surface forces acting between the foam film surfaces (such as van der Waals, electrostatic, depletion, etc. [17,18]). For simplicity, below we consider explicitly only van der Waals (vdW) attraction between film surfaces:

$$\Pi(h) = -A_H/6\pi h^3 \qquad (2)$$

where $A_H$ is Hamaker constant, which depends on the dielectric constants and refractive indexes of the dispersed and continuous phases [17,18].

It is well known from literature [14,19-23] that in presence of attractive forces, the thinning of the liquid films (foam or emulsion) down to a certain critical thickness, $h_{CR}$, leads to spontaneous jump to a very small thickness, often corresponding to two surfactant monolayers stabilizing the films by a short-range steric repulsion, see Fig. 2. This jump is driven by the attractive interactions and has been studied in detail both theoretically and experimentally in relation to coalescence stability of foams and emulsions [19,21-22]. The formed ultrathin films, called "black films" (BF) because they appear very dark when observed in reflected light, are characterized by strong attraction between the film surfaces and are thus able to withstand a certain detachment force [14,20,22]. Therefore, the formation of BF leads to adhesion between the neighboring bubbles (drops), which could jam locally the foam (emulsion).

The spontaneous jump in film thickness at $h_{CR}$ is driven by attractive surface forces and the following explicit formula for $h_{CR}$ was derived for prevailing van der Waals interactions [19]:

$$h_{CR} = 0.21\left(\frac{A_H^2 R_F^2}{\sigma P_C}\right)^{1/7} \qquad (3)$$

The closed set of Eqs. (1)-(3) provides the theoretical basis for considering the effect of film thinning on foam and emulsion jamming in the case of prevailing vdW interactions (for other attractive interactions, the same approach can be used at known $\Pi(h)$). By comparing the thickness of the dynamic film, formed between two neighboring bubbles in sheared foam, $h(t)$, with the critical film thickness, $h_{CR}$, one can determine whether the foam film will spontaneously jump to form BF, while the bubbles pass by each other, thus inducing strong bubble-bubble adhesion. Such a comparison is illustrated in Fig. 2, with parameter values, chosen to be typical for bubbles in sheared foams.

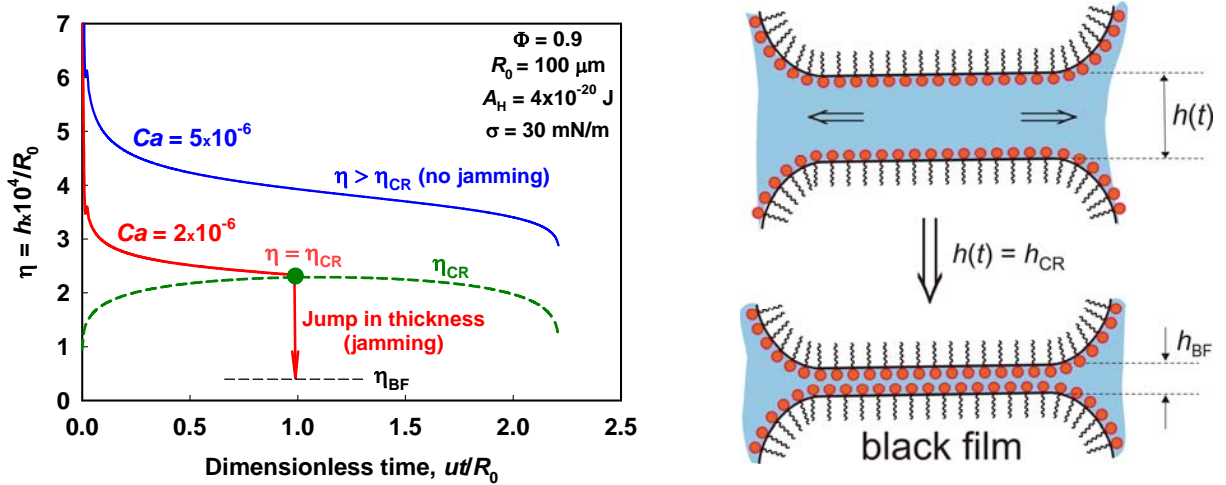

**FIG. 2.** (a) Calculated dimensionless thickness of the foam film between two neighboring bubbles in sheared foam, $\eta = h/R_0$, as a function of dimensionless time after film formation, $ut/R_0$, for two shear rates (solid curves), compared to dimensionless critical film thickness, $\eta_{CR} = h_{CR}/R_0$ (dashed curve). The transient film thickness becomes equal to the critical thickness, $h(t) = h_{CR}$, only below a certain critical capillary number (shear rate). (b) Schematic presentation of the thinning of the foam film between two bubbles. After the spontaneous jump in film thickness at $h(t) = h_{CR}$, the formed very thin (black) film is stabilized by short-range repulsive forces, whereas the long-range attractive forces (e.g. van der Waals) lead to adhesion between film surfaces [17,18,20,22].

As seen from Fig. 2(a), $h(t) = h_{CR}$ when $R_F(t)$ in close to its maximum, viz. at dimensionless time $\tilde{t} = ut/R_0 \approx 1$ (confirmed with other numerical calculations as well). This observation allows us to derive relatively accurate approximate expressions for the critical shear rate, $\dot{\gamma}_{jam}$, and for the critical capillary number, $Ca_{jam}$, leading to spontaneous jump in the film thickness and, thereby, to foam jamming. With this aim in view, we first integrate Eq. (1) at negligible disjoining pressure, $\Pi(h) \ll P_C$, to obtain the following expression for the film thickness [11,12]:

$$\frac{1}{h^2} = \frac{1}{h_0^2} + \frac{16 P_C}{3\mu} \frac{1}{u\sqrt{4R_{EFF}^2 - l_m^2}} \left[ \text{ArcTanh}\left(\frac{\sqrt{l_0^2 - l_m^2}}{\sqrt{4R_{EFF}^2 - l_m^2}}\right) + \text{ArcTanh}\left(\frac{ut - \sqrt{l_0^2 - l_m^2}}{\sqrt{4R_{EFF}^2 - l_m^2}}\right) \right] \quad (4)$$

Note that the attractive van der Waals forces are neglected in Eq. (4) which is, therefore, only an approximation to the more precise result, which can be obtained straightforwardly by numerical integration of Eq. (1). Next, we estimate from Eq. (4) the film thickness in moment $\tilde{t} = 1$ (viz. at $l =$ $l_m$), taking into account the fact that the term with $h_0$ is usually negligible (because $h_0 \gg h_{CR}$) and the second term in the square brackets is identically zero in this moment. The result of this estimate is:

$$h(\tilde{t}=1) \approx 0.18(R_N R_0\, Ca)^{1/2} \qquad \text{(at negligible surface forces)} \quad (5)$$

Next we use the relation between $R_N$ and the capillary pressure of the bubbles, $R_N = 2\sigma/P_C$, to derive an approximate interpolating expression, $R_N \approx 1.86 R_0 (1-\Phi)^{0.4}$, from a known expression for $P_C$ [16] (this expression for $R_N$ is valid for $0.80 \leq \Phi \leq 0.98$). Finally, we note that the film radius at $\tilde{t}=1$, could be approximated as $R_F(\tilde{t}=1) \approx R_0/2$ for not-very-low volume fractions. Combining all these expressions with the condition for occurrence of film instability, $h(\tilde{t} \approx 1) = h_{CR}$, we derive the following explicit expressions for the critical shear rate and the critical capillary number leading to foam and emulsion jamming:

$$Ca_{jam} = \frac{0.43}{(1-\Phi)^{2/7}} \left(\frac{A_H}{\sigma R_0^2}\right)^{4/7} \qquad (6)$$

$$\dot{\gamma}_{jam} = 0.43 \frac{\sigma^{3/7} A_H^{4/7}}{\mu R_0^{15/7} (1-\Phi)^{2/7}} \qquad (7)$$

Equations (6) and (7) were compared with the numerical results from the integration of Eq (1), taking into account the correct dependence of $R_F$ on $\Phi$, and were found to give correct values within 10-15 % for bubble volume fractions $0.80 \leq \Phi \leq 0.98$ and capillary numbers, $10^{-8} < Ca_{jam} < 10^{-4}$, which are of primary interest. Therefore, Eqs. (6)-(7) are appropriate and convenient for illustrating the main trends and for simple estimates.

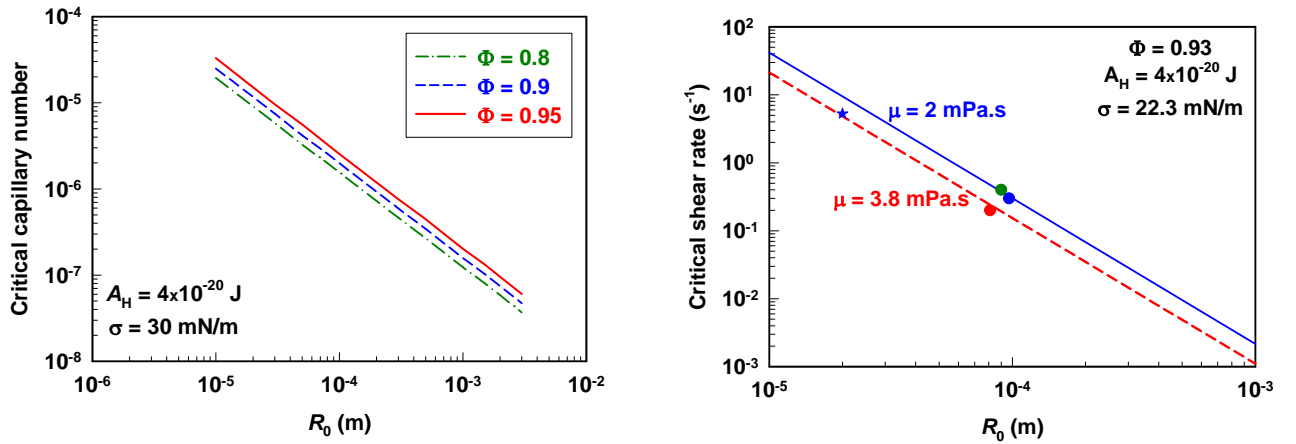

**FIG. 3.** (a) Numerical results for the dependence of $Ca_{jam}$ on bubble radius, $R_0$, at different bubble volume fractions, $\Phi$, with Hamaker constant, $A_H$, and interfacial tension, $\sigma$, taken typical for foams. (b) Comparison of our experimental data for $\dot{\gamma}_{jam}$, obtained with foams (circles) [24], with the theoretical prediction (lines). Asterisk shows data for shaving foam reported in [4].

Results from the numerical calculations for $\dot{\gamma}_{jam}$ and $Ca_{jam}$, with parameters typical for foams, are shown in Fig. 3(a). One sees that the dependence of $\dot{\gamma}_{jam}$ and $Ca_{jam}$ on bubble volume fraction is relatively weak, whereas the dependence on the bubble size (which varies in the real systems in a wide range) is significant. Similar numerical calculations showed that the dependence of $\dot{\gamma}_{jam}$ and $Ca_{jam}$ on Hamaker constant, $A_H$, is also relatively weak, when the latter is varied in the typical range for foams and emulsions (between $4\times10^{-21}$ and $4\times10^{-20}$ J [14]). All these results are in a very good agreement with the approximate expressions, Eqs. (6) and (7), which also evidence strong dependence on the particle size only.

To check directly these theoretical predictions, we performed experiments with foams sheared between the parallel plates of a rheometer Gemini (Malvern Instruments, UK). Sandpaper was glued on the plates to suppress foam-wall slip and the periphery of the sheared foam was video-recorded, using a long-focus magnifying lens. The gap between the plates was 3 mm, whereas the average bubble radius was $\approx$ 150 μm. For each experiment, the velocity of more than 100 bubbles (with different vertical locations across the gap) was measured to reconstruct the foam velocity profile.

The observations showed that, at high shear rate, the foam flow is homogeneous, with relative rate of the neighboring bubbles representing well the average shear rate. In contrast, at very low shear rate, the layers of bubbles close to the plates were jammed, and the foam flow was realized through formation of a "slip" plane in the middle of the gap – the bubbles around this plane were jumping to the next positions, thus allowing for a relative motion of the two jammed zones attached to the plates.

Most interesting was the intermediate range of shear rates, in which a central zone was formed, where the bubbles were flowing with constant shear rate, coexisting with two jammed zones attached to the plates. In line with [6] we found that the variation of the global shear rate in this intermediate regime leads to variation of the widths of these zones, whereas the shear rate in the middle zone remains practically constant (lever rule) and corresponds to the critical shear rate, $\dot{\gamma}_{jam}$. Thus we determined $\dot{\gamma}_{jam}$ for several foams with different Φ, μ, and mean bubble size. The obtained results are compared with model predictions in Fig. 3(b). One sees a very good agreement, without adjustable parameters used in the calculations [24]. Reasonably good agreement was found also with the results obtained in [4] with shaving foams: $\dot{\gamma}_{jam}$ = 5.3 s$^{-1}$ versus 9.3 s$^{-1}$ predicted theoretically ($R_{32}$ = 20 μm, Φ = 0.93, μ = 1.85 mPa.s, σ = 22 mN/m).

There is also a qualitative agreement with the results obtained with 2D-foams (measured $\dot{\gamma}_{jam}$ = 0.07 s$^{-1}$ for $R_{32}$ = 1.7 mm [8]) and sheared emulsions (measured $\dot{\gamma}_{jam}$ = 1 s$^{-1}$ for $R_{32}$ = 0.15 μm [10]) in the sense that (1) the critical shear rate increases significantly with the decrease of particle size, and (2) jamming was observed only in presence of significant attractive forces. Note that no quantitative agreement is expected for these data: For the large bubbles studied in [8], gravity is expected to accelerate film thinning, thus leading to higher experimental critical rates. On the other side, the extremely small size of the emulsion drops studied in [10] corresponds to very fast film thinning, see Eq. (1), so that the surface forces could not be neglected (while being unknown for this system, which precludes numerical calculations). Also, the high capillary pressure of these small drops and the fact that Φ ≈ 0.73 was around the sphere close-packing mean that the drops could behave like solid spheres before jamming in [10], i.e. Eq. (7) is not applicable. In other words, we assume that the explanation for the observed in [7,10] jamming of dispersions is qualitatively the same (film instability) but additional factors, not included in our model affect the rate of film thinning and the resulting critical rate of jamming.

In conclusion, we present a theoretical model, explaining jamming/unjamming transitions in sheared foams and concentrated emulsions by critical instability of the films formed between the neighboring bubbles and drops. This instability leads to spontaneous, jump-wise decrease of film thickness at a certain critical thickness, $h_{CR}$, with a subsequent formation of very thin "black" films (BF) which are characterized with strong adhesion between the film surfaces. For this film instability to occur, the film should have enough time to thin down to $h_{CR}$ during the period of particle contact. The latter requirement allows us to calculate the critical shear rate, $\dot{\gamma}_{jam}$ (and the corresponding capillary number $Ca_{jam}$), leading to formation of BF and dispersion jamming. The model predicts that $Ca_{jam}$ depends mostly on particle size and on magnitude of attractive forces. The model predictions agree well with experimental data without using adjustable parameters. This agreement proves that properly chosen foams, with bubble radius in the range 20-150 μm, are indeed very appropriate for quantitative investigation of jamming and the related phenomena, because the bubble dynamics could be described theoretically in great detail and could be observed directly with optical devices.

After simple modifications, the same approach could be applied to other types of dispersions, such as suspensions of soft particles (vesicles, microgel particles) or spherical solid particles [1-3,6]. For this purpose, appropriate material parameters and expressions for the surface forces should be implemented in the consideration.

This study is supported by Unilever GRC, Trumbull, CT, USA, and by project Rila-422/2008 with the Bulgarian Ministry of Education and Science. It is related also to the activity of COST P21 action of the EC "Physics of drops".